\documentstyle[aps,preprint,epsf]{revtex}
\begin{document}

\title{Parity Dependence of Nuclear Level Densities}

\author{ Y. Alhassid$^1$, G.F. Bertsch$^2$, S. Liu$^1$ and
H. Nakada$^3$}

\address{$^1$Center for Theoretical Physics, Sloane Physics
Laboratory,
Yale University, New Haven, Connecticut  06520, U.S.A. \\
$^2$Institute for Nuclear Theory and Department of
Physics, Seattle, Washington 98195\\
$^3$Department of Physics, Chiba University,
Inage, Chiba 263-8522, Japan}

\date{\today}

\maketitle

\begin{abstract}

A simple formula for the ratio of the number of odd- and
even-parity states as a function of temperature is derived. This
formula is used to calculate the ratio of level densities of
opposite parities as a function of excitation energy. We test
the formula  with quantum Monte Carlo shell model calculations
in the $(pf+g_{9/2})$-shell. The formula describes well the
transition from low excitation energies where a single parity
dominates to high excitations where the two densities are equal.

\end{abstract}


Parity is a fundamental property of nuclear levels, and its
statistical distribution is important for describing
parity-violating processes and neutron-capture reactions.
Most theoretical models for level densities are based on the
Fermi gas model \cite{BM69}. Shell corrections and correlations
due to residual interactions are  included empirically.
An empirical modification of the Fermi gas formula -- the
backshifted Bethe formula (BBF) -- was successful in
fitting many experimental level densities by adjusting both
the single-particle level density parameter and the
backshift parameter \cite{Dilg73}. Only limited
data are available for the parity dependence of level densities
since the neutron $p$-wave resonances are much weaker than the
$s$-wave resonances at low energies and more difficult
to measure.  Ericson
\cite{Ericson} argued that the excitation of a relatively
small number of single-particle levels with opposite parity
can lead to an equal number of even- and odd-parity many-particle
densities. The assumption of equal densities of opposite
parities in the neutron resonance region is commonly
accepted \cite{HM72} and used in the calculations of
neutron-capture rates for $s$ and $r$ processes in
nucleosynthesis \cite{RTK97}.
Yet various theoretical studies \cite{parity} as well as
analysis of experimental data \cite{Rao95} indicate that
level densities can  have a significant parity dependence.

Parity properties can in principle be
calculated within the interacting shell
model, the basic theory of nuclear structure. However the
calculation of level densities in
the shell model requires large model spaces that are often
beyond the reach of conventional diagonalization methods.
Such methods are presently limited to $A \alt 50$ \cite{fp9,fp11}
(in the $pf$-shell). Recently, quantum Monte Carlo methods
\cite{LJK93,ADK94} were used to calculate total and
parity-projected level densities  \cite{NA97} in the framework
of the interacting shell model. The methods were applied to
nuclei in the iron-to-germanium region using the complete
$(pf+g_{9/2})$-shell model space.
The total level densities were found to be in good agreement
with the experimental level densities, and significant parity
dependence was found for $A \alt 65$.

  The Monte Carlo calculations of the parity-projected
level densities accurately take into account shell effects
and correlations due to the residual two-body interactions,
but they are computationally intensive. In this paper
we derive a simple formula for calculating the ratio of
the number of odd- and even-parity states as a function of
temperature. The formula is applied to nuclei in the iron
region and compared with the Monte Carlo calculations.
It reproduces well the crossover from low temperatures,
where one parity dominates, to higher temperatures, where
both densities become equal. Using the BBF for the total
level density, the results of the model can be converted
to a ratio of parity-projected level densities at fixed
excitation energies.

The Monte Carlo approach is based on the
Hubbard-Stratonovich representation of the many-body
imaginary-time propagator,
$e^{-\beta  H}= \int D[\sigma] G(\sigma) U_\sigma$,  where
$G(\sigma)$ is a Gaussian weight and $U_\sigma$ is a one-body
propagator that describes non-interacting nucleons moving in
fluctuating time-dependent fields $\sigma(\tau)$. The canonical
thermal expectation value of an observable $O$ can be written
as $\langle O
\rangle_ A = \int D[\sigma] G(\sigma){\rm Tr}_ A(O
U_\sigma)/\int
D[\sigma]  G(\sigma){\rm Tr}_A U_\sigma$, where ${\rm Tr}_A$
denotes a trace in the subspace of $A$ particles~\cite{particle}.
The integrand
is easily calculated by matrix algebra in the single-particle
space, and the multi-dimensional integral over the $\sigma$
fields is evaluated by the Monte Carlo methods.

Parity-projected level densities were calculated in the Monte
Carlo method using the projectors $P_\pm = (1 \pm P)/2$, where
$P$ is the parity operator \cite{NA97}.  For even-even nuclei,
the odd-parity level density is found to have large 
statistical Monte Carlo errors at lower energies 
(even for good-sign interactions) because of a sign problem
introduced by the projection on odd-parity states.  However,
the sign problem does not affect the odd-even ratio
of partition functions with the estimator
\begin{equation}\label{ratio-MC}
{Z_-\over Z_+} = \left.\left[ 1 - \left\langle
{\zeta_P(\sigma) \over \zeta(\sigma)} \right\rangle_W \right]
\right/\left[ 1 + \left\langle
{\zeta_P(\sigma) \over \zeta(\sigma)} \right\rangle_W \right]\;.
\end{equation}
Here $\zeta(\sigma) = {\rm Tr}_A U_\sigma$ and $\zeta_P(\sigma)
= {\rm Tr}_A (P U_\sigma)$. In (\ref{ratio-MC}) we have
used the notation
$\langle X_\sigma \rangle_W \equiv {\int D[\sigma]
W(\sigma) X_\sigma /
\int  D[\sigma] W(\sigma)}$,
where $W(\sigma)\equiv G(\sigma) {\rm Tr}_A U_\sigma$.
For a good-sign interaction and an even-even nucleus, $W$
is positive definite. In the Monte Carlo method we sample
the fields $\sigma$ according to $W(\sigma)$ and then
estimate $\langle  X_\sigma \rangle_W \approx
\sum_k X_{\sigma_k} /M$, where $\sigma_k$ are $M$ samples
of the fields.
At low temperatures $\langle \zeta_P(\sigma) /
\zeta(\sigma) \rangle_W \sim 1$, giving rise to a sign problem
for the odd-parity states.
The statistical Monte Carlo errors of $Z_\pm/Z$ are strongly
correlated (since $Z_-/Z + Z_+/Z=1$) and the error estimate
of the ratio in (\ref{ratio-MC}) is $\Delta \left({Z_-/Z_+}
\right)  \approx
\left({Z_+/ Z} \right)^{-2}\Delta \left( {Z_+ / Z} \right)$.

\begin{figure}

\vspace{2 cm}

\epsfxsize= 15 cm\centerline{\epsffile{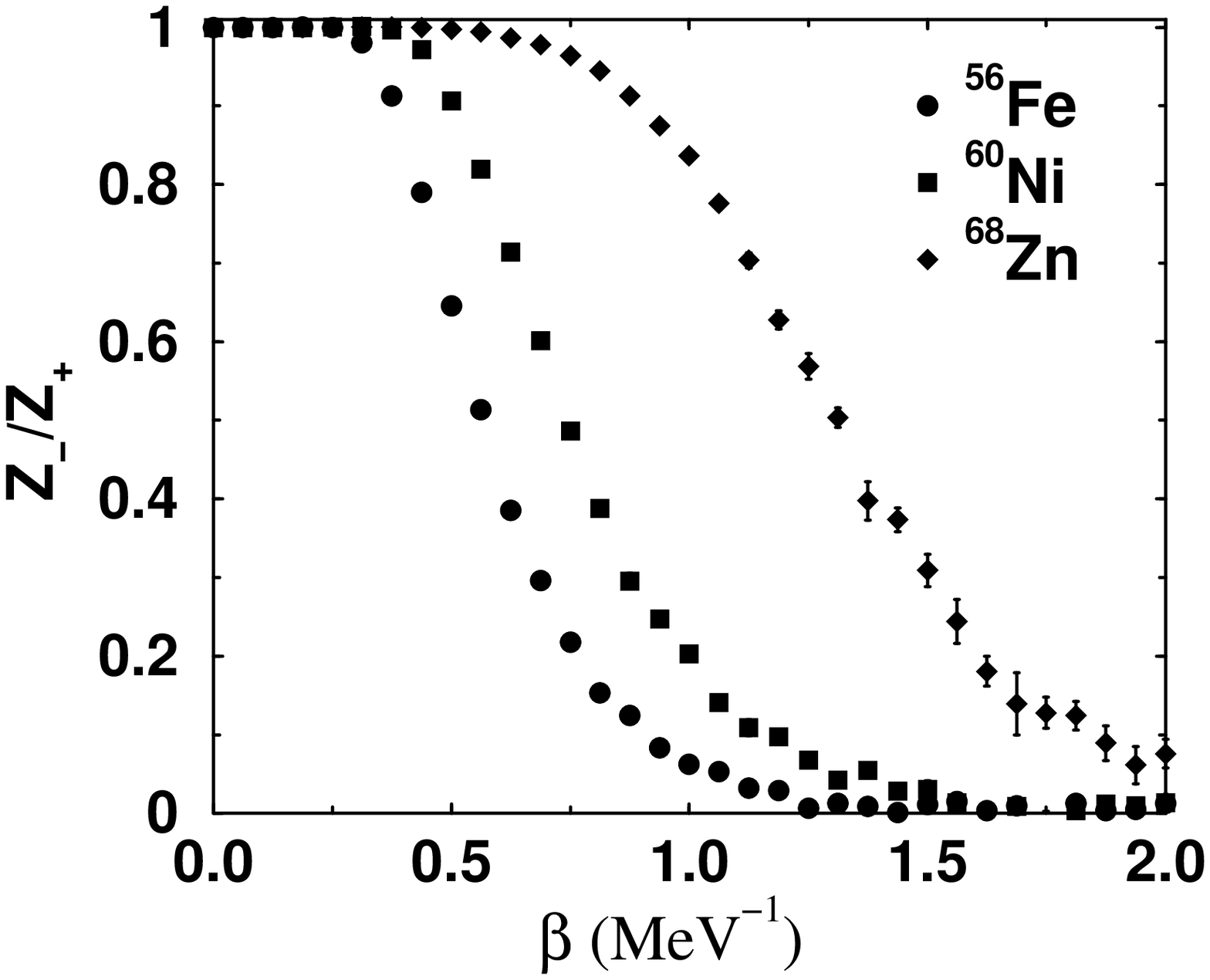}}

\vspace{1 cm}

\caption { Ratio of odd- to even-parity partition functions
$Z_-(\beta)/Z_+(\beta)$ versus inverse temperature
$\beta$ for $^{56}$Fe (circles), $^{60}$Ni (squares)
and $^{68}$Zn (diamonds), calculated in the Monte Carlo method
according to Eq. (\protect\ref{ratio-MC}).}
\label{fig:ratio-MC}

\newpage

\vspace*{2 cm}

 \epsfxsize= 14 cm \centerline{\epsffile{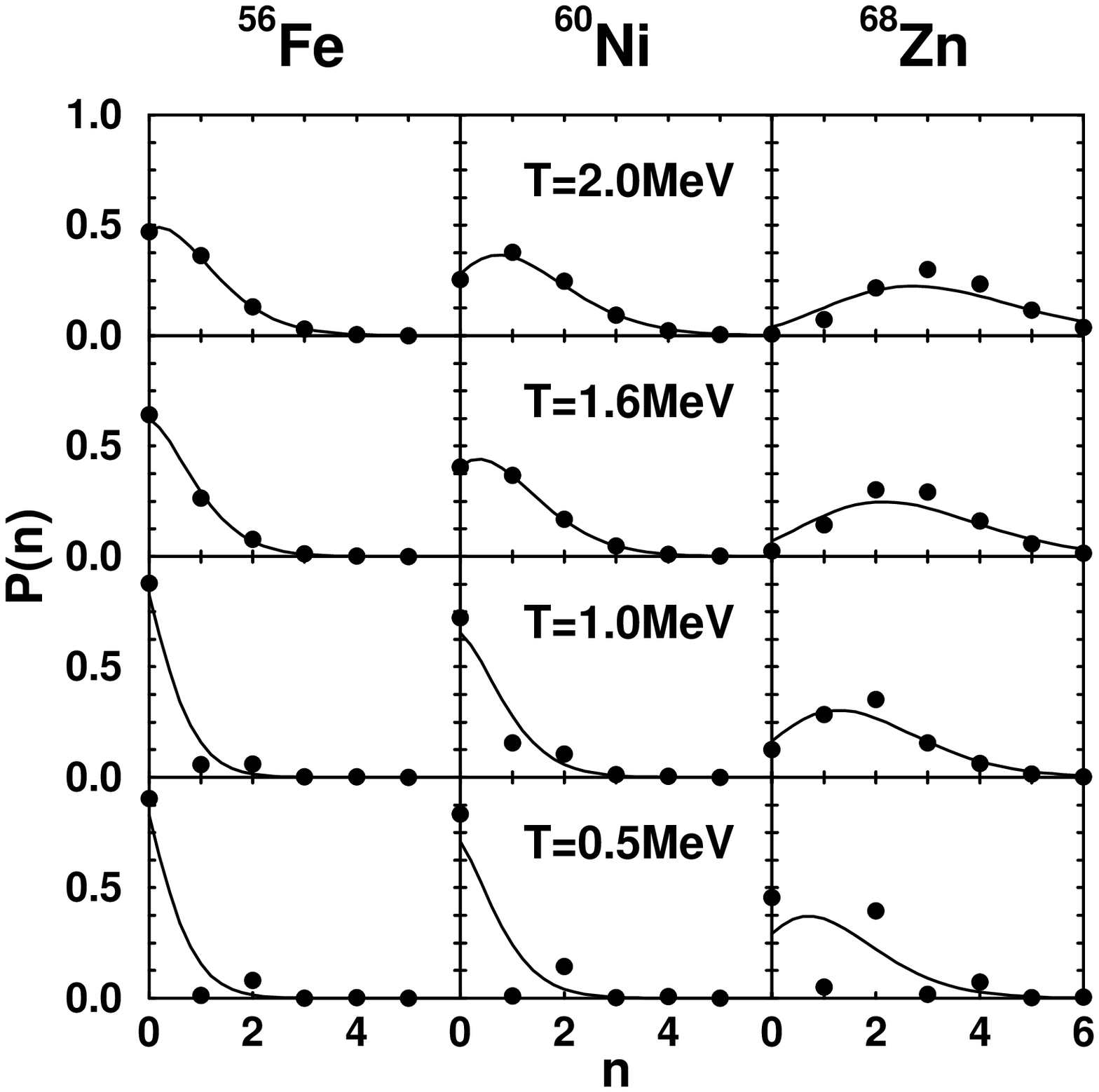}}

\vspace{1 cm}

\caption { The distribution $P(n)$ of the number of particles
$n$ occupying the even-parity $g_{9/2}$ orbitals. The solid
circles are the Monte Carlo results using
(\protect\ref{n-dist}). The solid lines are Poisson
distributions (\protect\ref{Poisson}) with the same values of
$\langle n \rangle$ calculated in the Monte Carlo method.}
\label{fig:Poisson}

\newpage

\vspace*{2 cm}

\epsfxsize=15 cm \centerline{\epsffile{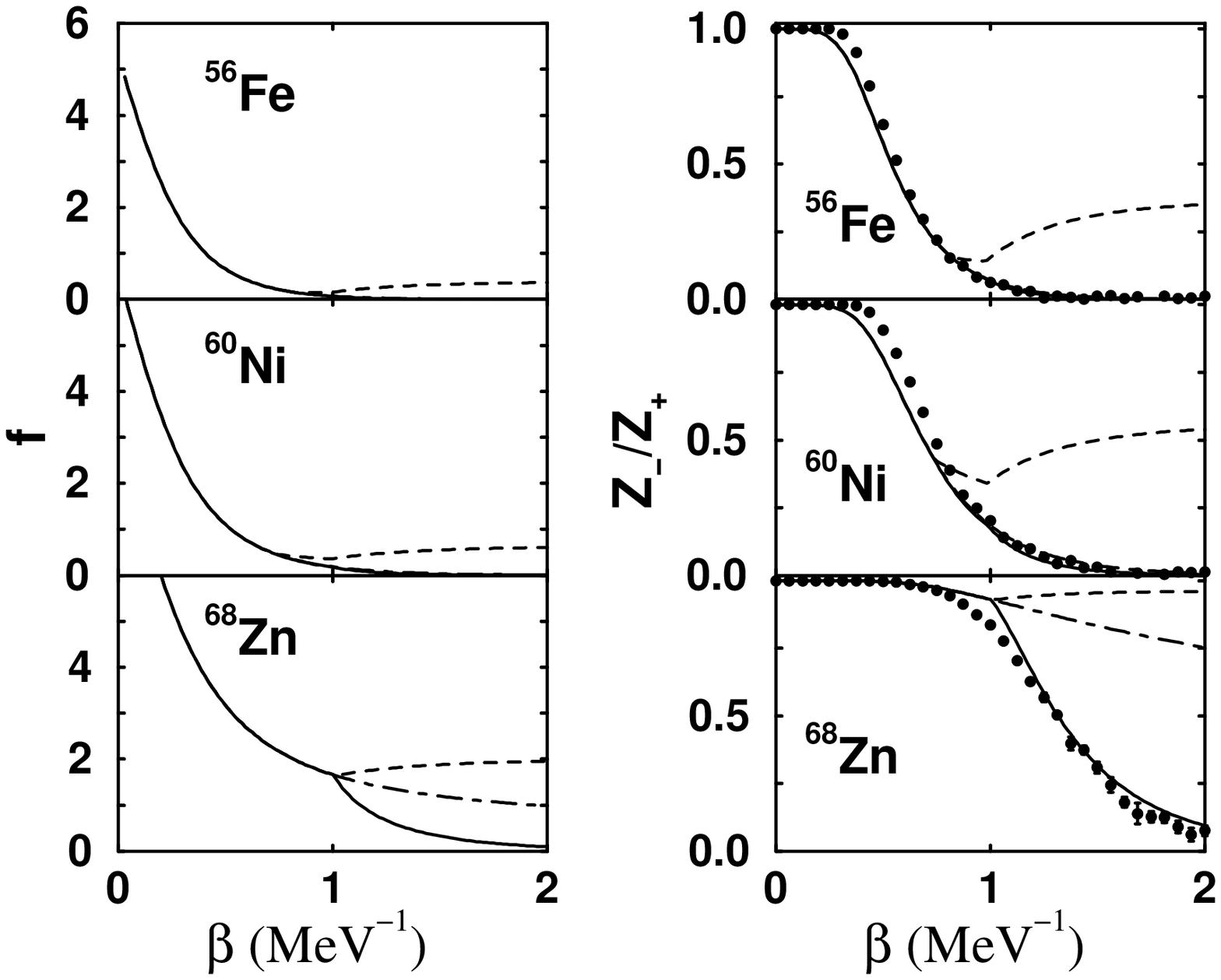}}

\vspace{1 cm}

\caption {Left column: The solid lines are the occupation $f$ of
the quasi-particle even-parity levels
(\protect\ref{f-occupation}) versus  $\beta$ for $^{56}$Fe,
$^{60}$Ni and $^{68}$Zn. Above the BCS temperature $f$ coincides
with the average occupation $\langle n\rangle$ of the
even-parity states.  The dashed lines show $\langle n \rangle$
below the BCS temperature (Eq. (\protect\ref{n-occupation}))
while the dotted-dashed lines are calculated from the
Fermi-Dirac occupations. Right column: the ratio $Z_-/Z_+$
versus $\beta$ calculated from (\protect\ref{parity-ratio})
using  the occupations $f$ shown on the left (solid, dashed and
dotted-dashed lines). For comparison we show by symbols the
Monte Carlo results. } \label{fig:z-ratio}

\newpage

\vspace*{2 cm}

\epsfxsize=12 cm \centerline{\epsffile{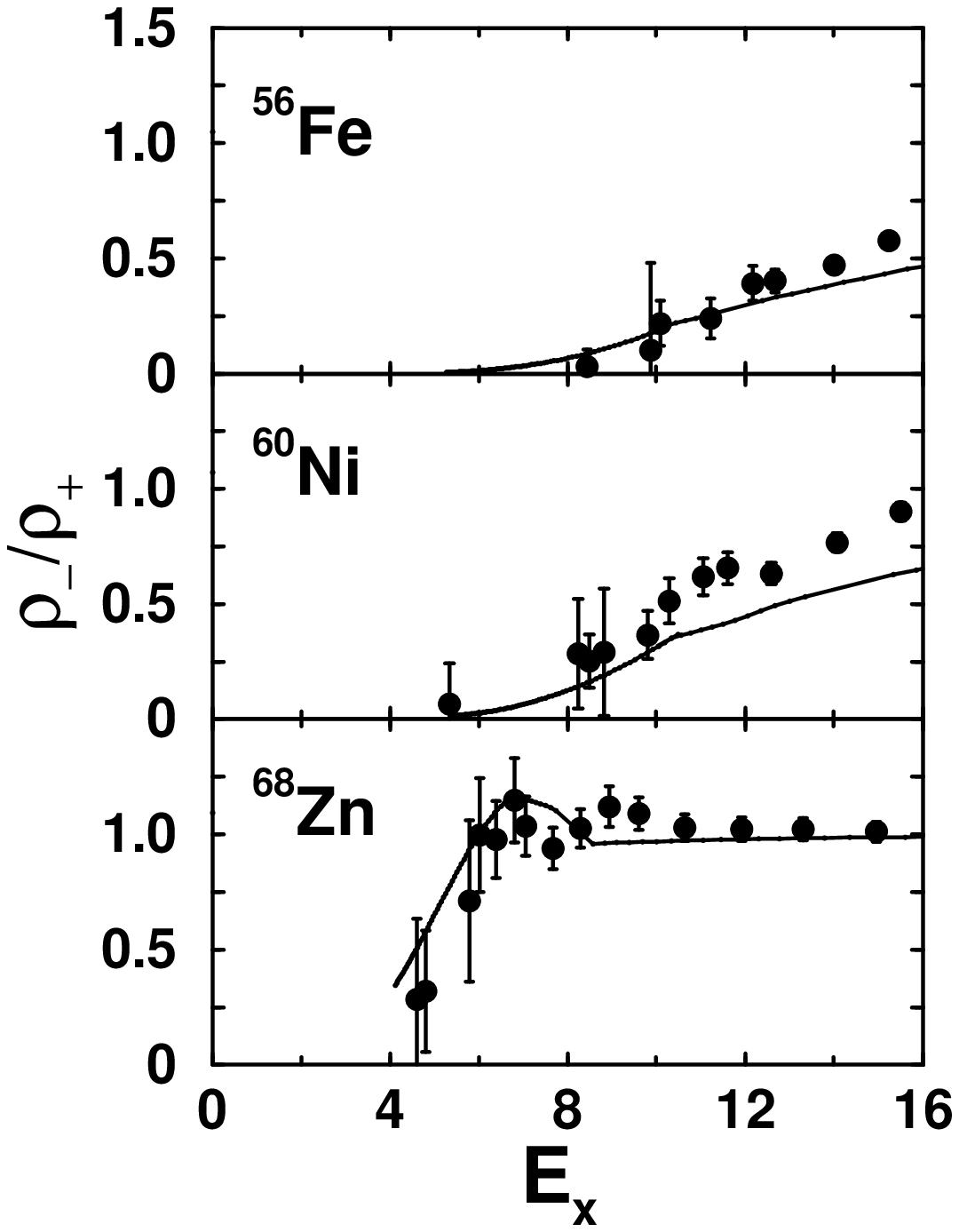}}

\vspace{1 cm}

\caption
{The parity ratio $\rho_-(E_x)/\rho_+(E_x)$ versus
excitation energy $E_x$. The solid lines are calculated
 from Eqs. (\protect\ref{parity-ratio}) and
(\ref{f-occupation}) (see text), and the symbols with error
bars are obtained
in the Monte Carlo method of Ref. \protect\cite{NA97}
(shown with error bars).
}
\label{fig:level-ratio}

\end{figure}

We have calculated the ratio $Z_-/Z_+$ for nuclei in the iron
region using the complete $(pf+g_{9/2})$-shell and the good-sign
interaction of Ref. \cite{NA97}. This interaction properly
 includes the dominant collective components of realistic 
effective nuclear interactions \cite{DZ96}.  
In Fig. \ref{fig:ratio-MC}
we show the ratio of odd- to even-parity states as a function
of inverse temperature $\beta$ for three nuclei in the iron
region: $^{56}$Fe, $^{60}$Ni and $^{68}$Zn. In all three
cases we observe a transition from mostly even-parity
states at low temperatures to an equal number of opposite
parity states
at high temperatures. However the crossover depends
on the nucleus. For example, the crossover occurs at lower
temperatures for $^{68}$Zn than for $^{56}$Fe.

  This observed parity dependence can be explained
quantitatively by a simple model.
 We divide the single-particle levels into two
groups of even- and odd-parities, and denote the group
having the smaller average occupation by  $\pi$.   If the
particles occupy the single-particle states independently
and randomly, we expect the distribution
of occupancies $n$ of the $\pi$ parity group to be Poisson:
\begin{equation}\label{Poisson}
P(n) = {f ^n \over n!} e^{-f}\;.
\end{equation}
Here $f$ is the average occupancy of orbitals
with parity $\pi$, which depends on temperature. We have assumed
that the total number of the particles is sufficiently large
compared with $f$.

For a nucleus with even $A$, an odd-parity many-particle
state is obtained when $n$ is odd. The probability to have
an odd-parity state is thus $P_- = \sum_n^{\rm odd}
P(n) = e^{-f} \sinh f$, while the probability for an
even-parity state is $P_+ = \sum_n^{\rm even}
P(n) = e^{-f} \cosh f$. The ratio is then
\begin{equation}\label{parity-ratio}
{P_- \over P_+} = {Z_-(\beta) \over Z_+(\beta)} = \tanh f \;,
\end{equation}
where we have identified $P_{\pm} =Z_\pm/Z$ in terms
of  the partition functions $Z_\pm(\beta)$ of the
even/odd parity states and the total partition
function $Z(\beta)$.

The argument leading to Eq. (\ref{parity-ratio}) is easily
extended to the case where protons and neutrons are treated
separately.  We denote by $P_p(n_p)$ and $P_n(n_n)$ the
respective Poisson distributions for protons and neutrons
with average occupancies $f_p$ and $f_n$ of single-particle
states with parity $\pi$.  A many-particle odd-parity
state results if protons and neutrons have overall parities
$(+,-)$ or $(-,+)$, respectively, i.e.,   $P_-= P_{p +}
P_{n -} + P_{p -} P_{n +}= e^{-(f_p+f_n)} \left(\cosh f_p
\sinh f_n + \sinh f_p \cosh f_n \right)$ (for an even-even
nucleus). Similarly we find $P_+ = P_{p +} P_{n +} + P_{p -}
P_{n -}= e^{-(f_p+f_n)} \left(\cosh f_p \cosh f_n + \sinh f_p
\sinh f_n \right)$. The parity ratio is then given by Eq.
(\ref{parity-ratio}) but with $f = f_p +f_n$. Furthermore,
 the convoluted distribution $P(n)= \sum\limits_{n_p+n_n=n}
 P_p(n_p)P_n(n_n)$ of finding $n = n_p+n_n$ nucleons in
 orbitals with parity $\pi$ is by itself a Poisson distribution
with $f=f_p+f_n$.

Above the pairing transition temperature and in the 
independent particle model,
$f=\langle n\rangle$ is evaluated from the
Fermi-Dirac distribution
$\langle n \rangle = \sum_{a \in \pi} \{ 1+
\exp[\beta(\epsilon_a - \mu)]\}^{-1}$,
where the sum is over all orbitals of parity $\pi$, and
the chemical potential $\mu$ is determined from the total
number of particles (in practice we use different chemical
potentials for protons and for neutrons). To mimic
interaction effects we use a single-particle spectrum that
corresponds to an axially deformed Woods-Saxon potential.
For axial deformations, parity is a good quantum number,
and we can estimate $\langle n\rangle$ by summing over the
Fermi-Dirac occupations of all deformed orbitals with parity $\pi$.

To test how well the Poisson distribution (\ref{Poisson})
describes the distribution $P(n)$ of the occupation of the
single-particle states with parity $\pi$, we  compare
with the Monte Carlo results.
The probability to find $n$ particles in single-particle states
with parity $\pi$ is
\begin{equation}\label{n-dist}
P(n) = {{\rm Tr}_A \left[e^{-\beta H} \delta(\hat n -n)\right]
\over {\rm Tr}_A e^{-\beta H}}
=\left\langle { {\rm Tr}_A \left[U_\sigma
\delta(\hat n - n)\right] \over {\rm Tr}_A U_\sigma}
\right\rangle_W \;.
\end{equation}
The quantity inside the brackets of Eq. (\ref{n-dist})
is calculated from a double projection on particle
number $A$ and occupation number $n$ of states with
parity $\pi$
\begin{equation}\label{n-project}
{\rm Tr}_A \left[U_\sigma \delta(\hat n - n)\right] =
{1 \over N(N_\pi+1)} \sum\limits_{m=1}^{N}
\sum\limits_{k=0}^{N_\pi}
e^{-i\phi_m A} e^{-i\varphi_k n} \det\left(1 + e^{i \phi_m}
e^{i \varphi_k I_\pi} {\bf U}_\sigma \right) \;,
\end{equation}
where $I_\pi$ is a diagonal matrix with diagonal elements
$1$ for each orbital of parity $\pi$ and $0$ otherwise.
The quadrature points are given by $\phi_m = 2\pi m /N$
and $\varphi_k = 2\pi k/(N_\pi+1)$, where $N$ is the total
number of single-particle states and $N_\pi$ is the number
of orbitals with parity $\pi$. ${\bf U}_\sigma$ is the $N
\times N$ matrix representing the propagator $U_\sigma$ in
the single-particle space.

For nuclei in the iron region the occupation of the
even-parity orbital $g_{9/2}$ is relatively small,
and we choose $n$ to denote the occupation of the $g_{9/2}$
states. Using the complete $(pf+g_{9/2})$-shell
we calculated the distributions $P(n)$ from Eqs. (\ref{n-dist})
and (\ref{n-project}). The results are shown in Fig.
\ref{fig:Poisson}
for $^{56}$Fe, $^{60}$Ni and $^{68}$Zn at several temperatures
(solid circles). The solid lines are the Poisson distributions
(\ref{Poisson}) with $f$ taken to be the average occupation of
the $g_{9/2}$ orbital (calculated in the Monte Carlo). At
high temperatures the microscopic distributions are well
described  by the Poisson distribution. However, for
temperatures $T \alt 1$ MeV we observe deviations describing
the enhancement of $P(n)$ for even  $n$ and the suppression
for odd $n$ due to pairing effects.

At lower temperatures $T \alt 1$ MeV, it is necessary to take
into account pairing effects.  We can still use Eq.
(\ref{parity-ratio}), but now with quasi-particles.
Consequently, Eq. (\ref{Poisson}) is
applicable where $n$ is replaced by the number of
quasi-particles with parity $\pi$, and $f$ in Eq.
(\ref{parity-ratio}) is the average occupation of
quasi-particle states with parity $\pi$.

As in the original BCS treatment \cite{BCS}, the
occupation probabilities have a component $v_a^2$
from condensed pairs  and
a component $f_a$ from quasi-particles.
Minimizing the free energy gives the quasi-particle occupation
factor
$f_a = 1 / [1+ \exp(\beta E_a)]$,
where $E_a=\sqrt{(\epsilon_a-\lambda)^2 + \Delta^2}$ and the
gap $\Delta$ and the chemical potential $\lambda$ (at
finite $T$) satisfy
self-consistency conditions. Since the condensed pair
occupations play no role in the parity of the states,
one should only use the quasi-particle $f$ in Eq.
(\ref{parity-ratio})
\begin{equation}\label{f-occupation}
f = \sum_{a \in \pi} f_a = \sum_{a \in \pi}
{ 1\over 1+ \exp(\beta E_a)} \;.
\end{equation}

We have applied the above model to determine the parity-ratio
  of levels for the three nuclei shown in Fig. \ref{ratio-MC}.
The deformation parameter $\delta$ (used to calculate
the single-particle spectrum $\epsilon_a$) is extracted
from the experimental $B(E2)$ values \cite{BE2}
for the $2^+ \to 0^+$
transition using $B(E2) = [(3/4\pi) Ze r_0^2 A^{2/3} \delta]^2/5$.
We find (using $r_0=1.27$ fm) $\delta= 0.22, 0.18$ and
$0.19$ for $^{56}$Fe, $^{60}$Ni and $^{68}$Zn, respectively.
The pairing gap $\Delta$ (at $T=0$) is extracted from
odd-even mass differences and is used to determine the
pairing strength $G$ by a BCS calculation.   The total
occupation $f$ of the quasi-particle even-parity
states (\ref{f-occupation}) is shown by the solid lines
on the left column of Fig. \ref{fig:z-ratio}. Above
the critical BCS temperature $f$ coincides with
$\langle n \rangle$. Below the BCS temperature the average
number of particles with parity $\pi$ is different
from $f$ and includes a contribution from condensed pairs
\begin{equation}\label{n-occupation}
\langle n\rangle =  \sum\limits_{a \in \pi}
\left[ f_a + v_a^2(1-2f_a)\right] =
{1\over 2} \sum\limits_{a \in \pi} \left[
1 -{\epsilon_a-\lambda\over E_a} \tanh \left({\beta E_a \over
2}\right)\right] \;.
\end{equation}
The calculated $\langle n \rangle$ is shown by the dashed
lines in Fig. \ref{fig:z-ratio}.  The parity-ratio $Z_-/Z_+$
is calculated from Eqs. (\ref{parity-ratio}) and
(\ref{f-occupation})  and shown by the solid lines on the
right column of Fig. \ref{fig:z-ratio}. The model 
describes well the Monte Carlo results (symbols).

Eq. (\ref{parity-ratio}) expresses the ratio
of the number of odd- and even-parity levels at
constant temperature.
To calculate the odd-to-even level density ratio at
constant excitation energy we use in addition the Lang
and  LeCouteur version \cite{LLC54} of the BBF for the
total level density $\rho(E_x)$.  We calculate
the total partition function from
$Z(\beta)= e^{-\beta E_{\rm gs}}
\int \rho(E_x) e^{-\beta E_x} d E_x$,
where $E_{gs}$ is the ground state energy.
Using $Z_+ + Z_-=Z$ and Eq. (\ref{parity-ratio}) for $Z_-/Z_+$,
we can determine $Z_\pm(\beta)= Z/(1 +\tanh^{\pm 1}f)$
and calculate the thermal energies for even- and odd-parity
states from $E_\pm=-\partial \ln Z_\pm/\partial
\beta= E + [d \ln (1 + \tanh^{\pm 1}f)/df](df/d\beta)$.
We can then calculate the canonical entropies and heat
capacities from standard thermodynamic relations and
find the parity-projected level densities $\rho_\pm(E_x)$.
   Fig. \ref{fig:level-ratio} shows the calculated ratio
$\rho_-(E_x)/\rho_+(E_x)$ in our model versus excitation
energy  (solid lines) for $^{56}$Fe, $^{60}$Ni and $^{68}$Zn.
The results compare well with the Monte Carlo calculations
shown by error bars.

In conclusion, we have derived a simple formula for the
parity dependence of level densities. The formula describes
the  crossover from low excitation energies where a single
parity dominates to higher excitations where odd- and even-parity
states have equal densities, and agrees well with
shell model Monte Carlo calculations.

This work was supported in part by the Department of
Energy grants No.\ DE-FG-0291-ER-40608 and DE-FG-06-90ER-40546,
and by the Ministry of Education, Science, Sports
and  Culture of Japan (grant 11740137).
Computational cycles
were provided  by the San Diego
Supercomputer Center (using NPACI resources), and by
the NERSC high performance computing facility at LBL.

\begin{thebibliography}{99}
\bibitem{particle}  For the simplicity of the discussion 
we assume here only one type of particles. In the
calculations we have treated separately protons and neutrons.
\end{thebibliography}
\end{document}